\documentclass[prb,twocolumn,aps,showpacs,10pt,floatfix,longbibliography]{revtex4-1}

\usepackage[pdftex]{graphicx} 
\usepackage{epstopdf}
\usepackage{verbatim}
\usepackage{color}
\usepackage{subfigure}
\usepackage{tabularx}
\usepackage{amsfonts}
\usepackage{wasysym}
\usepackage{bm}

\newcommand{\be}{\begin{equation}}
\newcommand{\ee}{\end{equation}}

\newcolumntype{C}[1]{>{\centering\let\newline\\\arraybackslash\hspace{0pt}}m{#1}}

\begin{document}
\title{Quantum spin ice on the breathing pyrochlore lattice}
\author{Lucile Savary$^2$}
\author{Xiaoqun Wang$^{3,6}$}
\author{Hae-Young Kee$^{4}$}
\author{Yong Baek Kim$^{4,5}$}
\author{Yue Yu$^{1,6}$}
\author{Gang Chen$^{1,6}$}
\email{gchen$_$physics@fudan.edu.cn}
\affiliation{${}^1$State Key Laboratory of Surface Physics, Center for Field Theory and Particle Physics, 
Department of Physics, Fudan University, Shanghai 200433, People's Republic of China}
\affiliation{${}^2$Department of Physics, Massachusetts Institute of Technology, 77 Massachusetts Ave., Cambridge, MA 02139}
\affiliation{${}^3$Department of Physics and astronomy, Shanghai Jiao Tong University, 
Shanghai 200240, Peoples Republic of China}
\affiliation{${}^4$Department of Physics, University of Toronto, 
Canadian Institute for Advanced Research/Quantum Materials Program, Toronto, Ontario MSG 1Z8, Canada}
\affiliation{${}^5$School of Physics, Korea Institute for Advanced Study, Seoul 130-722, Korea}
\affiliation{${}^6$Collaborative Innovation Center of Advanced Microstructures,
Nanjing, 210093, People's Republic of China}
\date{\today}
 
\begin{abstract}
The Coulombic quantum spin liquid in quantum spin ice is an exotic
quantum phase of matter that emerges on the pyrochlore lattice and is
currently actively searched for. Motivated by recent experiments on the Yb-based breathing pyrochlore
material Ba$_3$Yb$_2$Zn$_5$O$_{11}$, we 
theoretically study the phase diagram and magnetic properties of the relevant spin 
model. The latter takes the form of a quantum spin ice
Hamiltonian on a breathing pyrochlore lattice, and we analyze the stability of the 
quantum spin liquid phase in 
the absence of the inversion symmetry which the lattice breaks explicitly at lattice
sites. Using a gauge mean-field approach, we show that the quantum
spin liquid occupies a finite region in parameter space. Moreover, 
there exists a direct quantum phase transition between the quantum
spin liquid phase and featureless paramagnets, 
even though none of theses phases break any symmetry. 
At nonzero temperature, we show that breathing pyrochlores 
provide a much broader 
finite temperature spin liquid regime than their regular counterparts. 
We discuss the implications of the results for 
current experiments and make predictions for future experiments on breathing 
pyrochlores. 
\end{abstract}

\maketitle

\section{Introduction}
\label{sec1}

Frustrated magnetic materials provide a fertile arena to look for novel 
quantum phenomena. Frustration often leads to a large classical 
ground state degeneracy and quantum fluctuations are accordingly enhanced 
in quantum spin systems~\cite{Moessner2001,Balents10}. When strong quantum
fluctuations are taken to the extreme, they suppress any conventional 
magnetic order and may drive systems into a completely disordered quantum 
mechanical state, namely, a quantum spin liquid (QSL)~\cite{Lee05092008,Balents10}.  
QSLs are exotic quantum phases of matter, with long-range quantum entanglement, 
and are characterized by emergent gauge fields and deconfined fractionalized
excitations~\cite{Lee05092008,Balents10,Wenbook}.  

\begin{figure}[ht]
\centering
{\includegraphics[width=5.9cm]{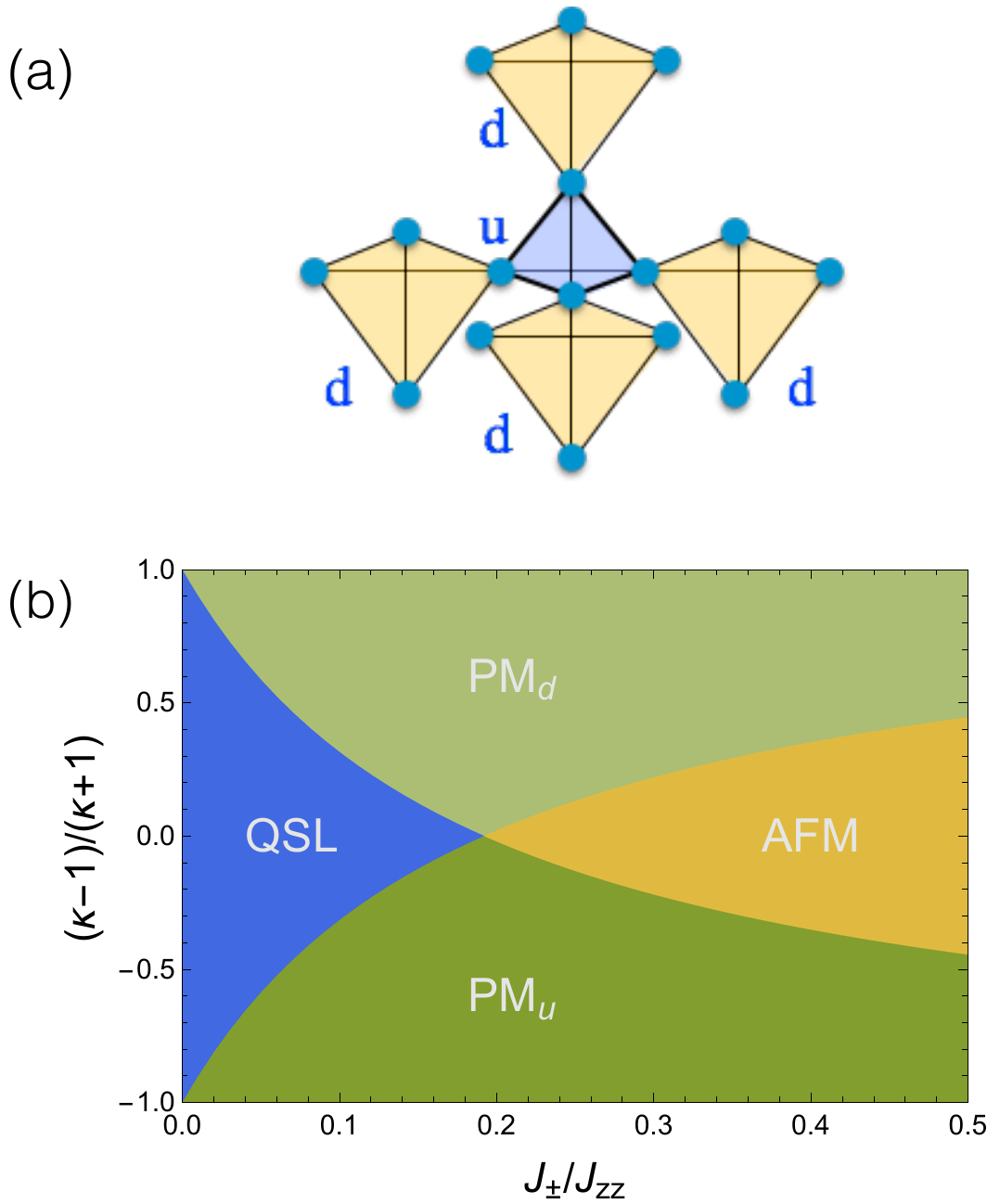}}
\caption{(Color online.) (a) The breathing pyrochlore lattice.  
The letter `u' (`d') refers to the up- (down-) pointing tetrahedra. 
The bond lengths on up and down tetrahedra are different. 
 (b) The phase diagram of the Hamiltonian $H$ in Eq.~(\ref{eq1}) at $T=0$.  
See the main text for a detailed discussion. }
\label{fig1}
\end{figure}

A three-dimensional $U(1)$ QSL with an emergent $U(1)$ gauge field 
and deconfined bosonic spinons has been proposed for rare-earth 
pyrochlore materials~\cite{Hermele04,PhysRevLett.98.157204,Ross11,Savary12,Sungbin12,Shannon12,Chen13,Gingras14,
Kimura_Nakatusji_2013,PhysRevLett.108.247210,PhysRevB.90.214430,li2016octupolar,Fritsch13,Zhou08,Chen2016}. 
In the context of rare-earth pyrochlores, Hamiltonians susceptible of
hosting such QSLs are often referred to as ``quantum spin ice'' (QSI)~\cite{PhysRevLett.98.157204,Ross11,Savary12,Sungbin12,Chen13,Gingras14},
and their QSL is called QSI QSL or Coulombic QSL. 
Despite intense theoretical and experimental efforts, 
no {\sl definitive} experimental evidence of 
such a QSL in QSI has been identified yet. 
This is partly because the underlying spin moment comes from 4f 
electrons whose exchange energy scale is usually very small. 
Thus, very challenging low temperature experiments are 
required to observe the intrinsic quantum properties of these materials,
such as spinon deconfinement and the emergent gapless gauge photon. 
This raises an important question: 
Can we find an alternative physical system that realizes the QSI QSL ground state 
at a higher energy scale? A proposal has suggested 
to replace the spin degrees of freedom 
by charge degrees of freedom. The resulting 
cluster Mott insulator on the pyrochlore lattice can realize QSI physics 
in the charge sector at a much higher temperature scale~\cite{Chen14a}. 
Recently, a new material, Ba$_3$Yb$_2$Zn$_5$O$_{11}$, 
where the Yb atoms form a breathing pyrochlore lattice 
(see Fig.~\ref{fig1}a), was synthesized. 
In contrast to the regular pyrochlore lattice, 
the breathing pyrochlore lattice has 
its up-pointing and down-pointing tetrahedra expanded 
and contracted, respectively, and thus breaks 
the lattice inversion symmetry at each lattice 
site~\cite{Cr_breathe2013,Rau,Cr_breathe2015,Haku,PhysRevLett.80.2933,PhysRevB.61.1149,Hirokazu2001,FeiYeLi2016}. 
Despite the a priori dominant antiferromagnetic interactions 
with a Curie-Weiss temperature $\Theta_{\text{CW}} = -6.7$K, 
the Yb local moments in Ba$_3$Yb$_2$Zn$_5$O$_{11}$ 
remain disordered down to $0.38$K~\cite{Kimura14}.
Motivated by the latter experiments, in this paper we propose that 
breathing pyrochlore materials are a 
new place to search for QSI physics, and in particular its QSL. 
Moreover, we argue that, due to the {\sl unique} structure of 
breathing pyrochlores, the transition or crossover temperature from QSI to the
higher-temperature thermal spin liquid phase occurs in breathing
pyrochlores at a much {\sl higher} temperature than in their 
`regular' counterparts.

The breathing pyrochlore lattice harbors the same symmetries as the regular 
pyrochlore lattice except that it lacks the inversion symmetry centered 
at lattice {\sl sites}. The spin model that we propose takes care of the 
absence of inversion symmetry by allowing different exchange couplings
on the up and down tetrahedra. 
Near the symmetric limit, the system and the model are understood 
from the known results on the regular pyrochlores. 
There exists a QSI U(1) QSL phase in this limit. 
To understand the robustness of the QSL phase,   
we extend the gauge mean-field approach to access the parameter regime 
for the breathing pyrochlore system. As one observes in Fig.~\ref{fig1}b, 
the QSL phase covers a large parameter regime. In addition, 
two paramagnetic (PM) phases are present in the asymmetric regime. 
These PM states are well understood in the strong 
asymmetric regime where the system breaks into decoupled tetrahedra.  
Since the ground state of the decoupled tetrahedron for our spin model 
is a singlet state, the PM in the asymmetric regime is smoothly 
connected to the simple product state of these decoupled 
tetrahedral singlets. 

In the QSI $U(1)$ QSL phase, there are gapless gauge photon modes
and deconfined and fractionalized spinon excitations. 
These exotic excitations are absent in the PM phases. 
In the gauge mean-field theory framework,
the quantum phase transition from the QSL phase to the nearby PM phases is 
understood as the condensation of the spinons. The gauge photon picks 
up a mass due to the spinon condensation. This is essentially 
the Anderson-Higgs' phenomenon, but it occurs in
a system where the gauge field is emergent. 

We further explore the finite temperature properties of the QSL phase. 
Both the perturbative argument and the gauge mean-field theory
show that the onset temperature of the finite temperature QSL regime
can be much higher than the regular pyrochlore system. This is because
the asymmetric exchange couplings on the breathing pyrochlores allow
the spins to fluctuate more effectively and thus enhance the energy scale 
of the collective spin fluctuations. The gauge mean-field theory is 
used to obtain the finite temperature phase diagram of the spin model. 
The persistence of the QSL physics to the high temperature regime
gives a large room for the experimental confirmation.

The remaining parts of the paper are organized as follows.
In Sec.~\ref{sec2}, we propose a minimal spin model for the breathing pyrochlore systems.
In Sec.~\ref{sec3}, we implement the gauge mean-field mapping and obtain the full phase diagram
of the spin model. We explain all the four phases and elucidate the nature of the 
phase transitions between them. In Sec.~\ref{sec4}, we extend the gauge mean-field theory to 
finite temperatures and explore the properties of the system in the finite temperature regime. 
Finally, in Sec.~\ref{sec5}, we discuss the experimental consequences of the different phases. 
In the appendices, we provide the details of the derivation and the theoretical framework.

\section{Minimal model for breathing pyrochlores}
\label{sec2}

In Ba$_3$Yb$_2$Zn$_5$O$_{11}$, the 4f electrons of 
a Yb$^{3+}$ ion form a $J=7/2$ local moment due to the strong
spin-orbit coupling of 4f electrons. 
The crystalline electric fields further split the eight-fold degeneracy 
of the $J=7/2$ manifold and lead to an on-site ground state with two-fold
Kramers degeneracy. This local ground state doublet is separated from the 
excited doublets by a large crystal field energy gap of 
$\approx 500$K~\cite{Kimura14}. 
Since the Curie-Weiss temperature $\Theta_{\text{CW}} = -6.7$K
is much smaller than the crystal field energy gap, one then 
introduces a pseudospin-1/2 operator, $\boldsymbol{\tau}$, 
that operates within the local ground state doublet, 
to describe the low temperature magnetic properties of 
Ba$_3$Yb$_2$Zn$_5$O$_{11}$.  
Moreover, the strong crystal fields and spin-orbit coupling lead to a
very large on-site anisotropy, which naturally singles out the
local three-fold axis of the $D_{3d}$ point group, which corresponds
to a $\langle 111\rangle$ crystallographic direction. Like the regular
pyrochlore lattice, the local $\langle 111\rangle$ direction at a 
site is the direction that points 
into or out of the centers of the neighboring tetrahedra. 
We define the $z$ components of the pseudospins to be along 
their local $\langle 111\rangle$ axis (see Appendix.~\ref{asec1}).

We consider a minimal model for the pseudospins $\tau = 1/2$ with Hamiltonian
\begin{eqnarray}
H &=& \sum_{\langle ij \rangle \in \text{u}} {J_{zz} {\tau}^z_i \tau^z_j 
- J_{\pm} (\tau^+_i \tau^-_j + \tau^-_i \tau^+_j )}
\nonumber \\
& + &
\sum_{\langle ij \rangle \in \text{d}} \kappa {\big[ J_{zz} {\tau}^z_i \tau^z_j 
- J_{\pm} (\tau^+_i \tau^-_j + \tau^-_i \tau^+_j ) \big]},
\label{eq1}
\end{eqnarray}
where ${\tau^{\pm}_{i} = \tau^x_i \pm i \tau^y_i}$. 
The hallmark of breathing pyrochlore systems is the absence of 
inversion symmetry centered at lattice sites, and the parameter 
$\kappa$ captures this property. More precisely, 
it parametrizes the asymmetry of the spin interactions between 
the up-pointing (labelled by `u') and down-pointing (labelled by `d') 
tetrahedra of the breathing pyrochlore lattice. On a regular pyrochlore lattice, 
due to the presence of inversion symmetry, 
one has $\kappa =1$. In breathing pyrochlore systems, 
the asymmetry parameter $\kappa$ should generically deviate from 1. 

Even though the Hamiltonian $H$ is not the most general spin model 
obtained from a full space group symmetry analysis of the 
breathing pyrochlore lattice \cite{Ross11,Savary12}, 
the Hamiltonian $H$ is sufficient 
for the purpose of understanding the stability of the QSI 
QSL phase against the breathing distortion, as well as some additional features of
breathing pyrochlore systems. 
Due to the strong spin-orbit coupling of the Yb 4f electrons, 
the interactions between local moments should be both bond 
dependent and anisotropic in (effective) spin space~\cite{William14,PhysRevB.82.174440,PhysRevB.84.094420}, which thus introduces additional complications. 
Its study will be discussed in future work.   

The Hamiltonian $H$ has been studied both theoretically and numerically on the 
regular pyrochlore lattice, where $\kappa=1$~\cite{Hermele04,Banerjee08,Savary12,Lv15}. There, in the Ising limit with $J_{\pm} = 0$, 
the system favors the extensively-degenerate ``2-in 2-out'' spin ice
ground state manifold.  
A small and finite transverse coupling $J_{\pm}$ generates a 
six-spin ring exchange around the hexagonal plaquettes of the pyrochlore lattice
and allows the system to fluctuate quantum mechanically
within the ice manifold~\cite{Hermele04,Savary12,savaryreview,Gingras14,Sungbin12,Chen13,PhysRevB.90.214430}. 
By mapping the effective ring exchange Hamiltonian
in the perturbative regime ($J_{zz} \gg |J_{\pm}|$) onto a lattice gauge theory 
that operates within the spin ice manifold, it was argued that the system exhibits a 
$U(1)$ QSL phase, {\it i.e.}\ the QSI QSL state~\cite{Hermele04}. 
This theoretical result was later confirmed numerically by 
quantum Monte Carlo calculations in the regime with $J_{\pm} >0$~\cite{Banerjee08,Lv15,Kato15,PhysRevLett.108.067204}. 
In the breathing pyrochlore lattice, the
perturbative argument in the limit $J_{zz} \gg |J_{\pm}|$ remains valid at least when the 
parameter $\kappa$ does not strongly deviate from 1 (see Appendix.~\ref{asec2}). 
Moreover, in general, the QSI QSL state is a robust quantum phase 
of matter and is stable against any small local perturbation~\cite{Hermele04}. 
Both arguments confirm that the QSL state of QSI should be the ground state of 
the minimal model $H$ at least in the regime with 
$J_{zz} \gg |J_{\pm}|$ and $\kappa \approx 1$. 

\section{Ground state phase diagram} 
\label{sec3}

To investigate the stability of the QSI QSL state further and to obtain its proximate 
phases on the breathing pyrochlore lattice, we apply the recently 
developed non-perturbative slave-particle construction \cite{Savary12,Sungbin12}. 
For this purpose, we first enlarge the physical Hilbert space 
to make the spinon and $U(1)$ gauge field variables 
explicit, and we express the effective spin operators as 
\begin{eqnarray}
\tau^+_i &\equiv & \tau^+_{{\bf r},{\bf r} + {\bf e}_{\mu}} 
= \Phi^{\dagger}_{\bf r} \Phi^{\phantom\dagger}_{{\bf r} +{\bf e}_{\mu}} 
\mathsf{s}^+_{{\bf r},{\bf r}+{\bf e}_{\mu}}, 
\label{eq2}
\\
\tau^z_i &\equiv &\tau^z_{{\bf r},{\bf r}+{\bf e}_{\mu}} 
= \mathsf{s}^z_{{\bf r},{\bf r}+{\bf e}_{\mu}},
\label{eq3}
\end{eqnarray}
where ${\bf r}$ belongs to the u sublattice of the diamond 
lattice that is formed by the centers of the up-pointing tetrahedra in
the breathing pyrochlore lattice (see Fig.~\ref{fig1}), where 
the vectors ${\bf e}_{\mu}$ (with $\mu =1,2,3,4$)
connect ${\bf r}$ with the centers of the neighboring tetrahedra, 
and $i$ is the breathing pyrochlore lattice site that is shared by the
two tetrahedra that are centered at the positions 
${\bf r}$ and ${{\bf r} + {\bf e}_{\mu}}$. $\Phi^\dagger_{\bf r}$ 
($\Phi^{\phantom\dagger}_{\bf r}$) is the (bosonic) spinon creation 
(annihilation) operator at site ${\bf r}$, and $\mathsf{s}^z_{{\bf r}{\bf r}'}, 
\mathsf{s}^{\pm}_{{\bf r}{\bf r}'}$ are ``spin''-$1/2$ operators that 
act as $U(1)$ gauge fields. To preserve the physical spin Hilbert space, 
we further impose the constraint 
\begin{equation}
{Q}_{\bf r} = \eta_{\bf r} \sum_{\mu} 
\mathsf{s}^z_{{\bf r},{\bf r} + \eta_{\bf r} {\bf e}_{\mu} } ,
\end{equation}
where $\eta_{\bf r} = \pm 1$ for ${\bf r} \in \text{u/d}$ sublattice. 
The operator ${Q}_{\bf r}$ counts the number of spinons and satisfies 
$[\phi_{\bf r}, Q_{{\bf r}' } ] = i \delta_{{\bf r}{\bf r}'}$,
where $\phi_{\bf r}$ is a $2\pi$ periodic angular variable defined as 
$\Phi_{\bf r} \equiv e^{-i \phi_{\bf r}} $, with 
$\Phi^\dagger_{\bf r}\Phi^{\phantom\dagger}_{\bf r} =1$, by construction. 
In terms of these slave particle operators, the Hamiltonian $H$ is rewritten as 
\begin{eqnarray}
H &=& \sum_{{\bf r} \in \text{u}} \frac{J_{zz}}{2} Q_{\bf r}^2 
- \kappa J_{\pm} \sum_{{\bf r}\in \text{d}, \mu\neq\nu} \Phi^\dagger_{{\bf r} 
- {\bf e}_{\mu}}\Phi^{\phantom\dagger}_{{\bf r} -{\bf e}_{\nu}} 
\mathsf{s}^-_{{\bf r},{\bf r}-{\bf e}_{\mu}} \mathsf{s}^+_{{\bf r},{\bf r}-{\bf e}_{\nu}} 
\nonumber \\
&+& 
\sum_{{\bf r} \in \text{d}} \frac{\kappa J_{zz}}{2} Q_{\bf r}^2 -  
J_{\pm} \sum_{{\bf r}\in \text{u}, \mu\neq\nu} \Phi^\dagger_{{\bf r} 
+ {\bf e}_{\mu}}\Phi^{\phantom\dagger}_{{\bf r} +{\bf e}_{\nu}} 
\mathsf{s}^-_{{\bf r},{\bf r}+{\bf e}_{\mu}} \mathsf{s}^+_{{\bf r},{\bf r}+{\bf e}_{\nu}}. 
\nonumber\\
\label{eq4}
\end{eqnarray}
The above Hamiltonian describes the bosonic spinons ($\Phi_{\bf r}$) 
hopping on the dual diamond lattice in the background of a fluctuating 
$U(1)$ gauge field ($\mathsf{s}^{\pm}_{{\bf r}{\bf r}'}$). 
It is manifestly invariant under the local $U(1)$ gauge transformation
$\Phi_{\bf r} \rightarrow \Phi_{\bf r} e^{-i \chi_{\bf r}}, \mathsf{s}^{\pm}_{{\bf r}{\bf r}'}
 \rightarrow \mathsf{s}^{\pm}_{{\bf r}{\bf r}'} e^{\pm i (\chi_{{\bf r}'} - \chi_{\bf r})}$.

We apply gauge mean field theory (gMFT)~\cite{Savary12,Sungbin12,Savary13}
to analyze the phase diagram of $H$. 
To proceed, we first note that the (low-energy) ring exchange Hamiltonian 
that operates within the spin ice manifold favors 
a zero gauge flux for $J_{\pm} > 0$ (see Appendix.~\ref{asec2}). 
Therefore, we choose a mean-field ansatz where the spinons experience 
a zero gauge flux through each hexagon of the pyrochlore lattice. 
We decouple Eq.~(\ref{eq4}) into the spinon and gauge field sectors, 
and select a gauge such that the spinon hopping is uniform. 
With this gauge choice, the gMFT connects 
to and reproduces the results of the $J_{\pm}\ll J_{zz}, |\kappa-1|\ll 1$  
limit where the solution is known from perturbation theory and 
numerical calculations~\cite{Hermele04,Savary12,savaryreview,Gingras14,Sungbin12,Chen13,PhysRevB.90.214430,Banerjee08,Lv15,Kato15,PhysRevLett.108.067204}. 
The gMFT phase diagram is depicted in Fig.~\ref{fig1}b.  
The QSL covers a wide range of $\kappa$ in parameter space. 
When the spinons on the u and d sublattices of the diamond lattice 
are both gapped (and thus do not condense), the system is in the QSI QSL phase 
(see Table~\ref{tab2}) and the low energy properties, such as
algebraic correlations, are controlled by the gapless $U(1)$ gauge photon.

\begin{table}[t]
\begin{tabular}{C{0.9cm}C{1.6cm}C{1.6cm}C{1.1cm}C{1.1cm}C{1.1cm}}
\hline\hline
Phase & $\langle \Phi_{\bf r} \rangle, {\bf r} \in \text{u}$ & 
$\langle \Phi_{\bf r} \rangle, \bf r\in \text{d}$ & 
$\langle \mathsf{s}^{\pm} \rangle$ & $\langle \mathsf{s}^{z} \rangle$
  & $|\langle \vec{\mathsf{s}} \rangle|$
\\
QSL & $ = 0$ & $=0$ & $\neq 0$ & $= 0$ & $\neq0$
\\
AFM & $ \neq 0 $ & $\neq 0$ & $\neq 0$ & $=0$ & $\neq0$
\\
PM$_{\text u}$ & $\neq 0 $ & $=0$ & $\neq 0$ &  $=0$ & $\neq0$
\\
PM$_{\text d}$ & $= 0 $ & $\neq 0$ & $\neq 0$ &  $=0$ & $\neq0$
\\
TSL & $=0$ & $=0$ & $=0$ & $=0$ & $=0$
\\
\hline\hline
\end{tabular}
\caption{Different ground state phases with their order parameters in
  gMFT. QSL and TSL stand for the quantum spin ice quantum spin
  liquid, and thermal spin liquid, respectively.}
\label{tab2}
\end{table}

We treat the QSI QSL as the parent state and analyze the 
proximate phases obtained via an Anderson-Higgs transition. 
We first consider leaving the QSL by condensing one spinon flavor
(i.e.\ condensing the spinons on one diamond sublattice) 
while keeping the other spinon flavor uncondensed. This is certainly 
reasonable due to the absence of on-site inversion symmetry in the breathing 
pyrochlore lattice. For concreteness, we condense the spinons on 
the down sublattice with $\langle \Phi_{\bf r}\rangle \neq 0$ 
for ${\bf r} \in $ d sublattice. As a consequence, the $U(1)$ gauge field picks up 
a mass through the usual Anderson-Higgs mechanism. 
What is surprising is that the resulting state is not magnetically ordered. 
This is because the spinons on the u sublattice are not condensed.
According to the slave particle construction in Eq.~(\ref{eq2}), this proximate
state of the QSI QSL, obtained by condensing spinons on one sublattice but
not on the other, does not develop any transverse magnetic long range order
following 
\begin{equation}
\langle \tau^{\pm}_{{\bf r}{\bf r}'} \rangle
= \langle \Phi_{\bf r}^\dagger \rangle \langle
\Phi_{{\bf r}'}^{\phantom\dagger} \rangle 
\langle \mathsf{s}^{\pm}_{{\bf r}{\bf r}'} \rangle = 0.
\end{equation} 
This paramagnetic state preserves time reversal symmetry
and all the lattice symmetries of the breathing pyrochlore
lattice. We thus dub this paramagnetic (PM) 
state PM$_{\text u}$ in Fig.~\ref{fig1}b and Table~\ref{tab2}.

To understand the nature of the PM$_{\rm u/d}$ phases better,
especially outside of gMFT, we consider the limiting
case $\kappa \rightarrow 0$. In this special limit, 
the breathing pyrochlore system reduces to a set of decoupled 
up-pointing tetrahedra, i.e.\ set of four spins. Since a
tetrahedron contains a finite number of spins (it is a finite system),
its ground state must preserve all the symmetries of the system. For
each up-pointing tetrahedron, the local magnetization
$\sum_{i\in\text{u}}\tau^z_i$ commutes with the Hamiltonian and is
therefore a good quantum number, and the ground state 
is {\sl unique} with $\sum_{i\in\text{u}}\tau^z_i =0$. Since the tetrahedra
are decoupled, the {\sl many-body} ground state is
simply a product state of the single-up-tetrahedron ground states. Hence,
the system is an obvious paramagnetic state, preserving all symmetries.
A finite and small $\kappa$ introduces inter-up-tetrahedral 
couplings but does not significantly alter the ground state. 
Moreover, due to the weak $\tau^z$ coupling on the down-pointing tetrahedra,
the local magnetization ($\sum_{i\in\text{d}}\tau^z_i$) on each 
down-pointing tetrahedron is strongly fluctuating quantum mechanically
and thus is not a good quantum number. These features
are exactly reproduced by the gMFT description of the PM$_{\rm
  u}$: the spinon
condensate on the down tetrahedra corresponds to an ill-defined spinon
number $Q_{\mathbf{r}\in{\rm d}}=\sum_{i\in\text{d}}\tau^z_i$, and the
single-tetrahedron ground state with its (small) `cat' state on the up
tetrahedra preserves all symmetries and corresponds to
$\langle\Phi_{\mathbf{r}\in{\rm u}}\rangle=0$.  
Therefore, the gMFT approach provides a 
good description of the $\kappa \rightarrow 0$ limit, 
like it did for the QSI QSL phase.

The state obtained by instead condensing the spinons on the down sublattice, 
is naturally labelled PM$_{\text d}$. 
Like the relation between the PM$_{\text u}$ and
the $\kappa \rightarrow 0$ limit,  the 
PM$_{\text d}$ is smoothly connected to the paramagnetic state
in the limit $\kappa \rightarrow \infty$, and the gMFT 
also provides a good description of this limit. 
As the PM$_{\text u}$ and the PM$_{\text d}$ have neither 
gapless emergent gauge photon nor spontaneous continuous symmetry breaking, 
the system is fully gapped in these two phases. 
The quantum phase transition from the gapless QSI QSL to
the gapped PM$_{\text u}$ or PM$_{\text d}$ is described by the condensation
of one critical bosonic spinon mode coupled to a fluctuating $U(1)$ gauge field. 
This quantum phase transition is not present for the regular pyrochlores and is a new 
feature of the breathing pyrochlores.

A condensate of both spinon flavors is obtained for example by 
increasing $J_{\pm}/J_{zz}$ and keeping a moderate $\kappa$. 
The resulting state breaks time reversal symmetry and 
develops magnetic order via 
\begin{equation}
\langle \tau^{\pm}_{{\bf r}{\bf r}'} \rangle
= \langle \Phi_{\bf r}^\dagger \rangle \langle
\Phi_{{\bf r}'}^{\phantom\dagger} \rangle 
\langle \mathsf{s}^{\pm}_{{\bf r}{\bf r}'} \rangle \neq 0. 
\end{equation}
In terms of the original physical magnetic moments, 
this ordered state is an AFM state and the magnetic unit cell 
is identical to that of the crystal.  Moreover, the 
direct treatment of the limiting case with a dominant
$J_{\pm}/J_{zz}$ and a moderate $\kappa$ also leads
to a transverse spin ordering, which is again consistent with the 
results from the gMFT approach. 

\begin{figure}[tp]
\centering
 \includegraphics[width=7.5cm]{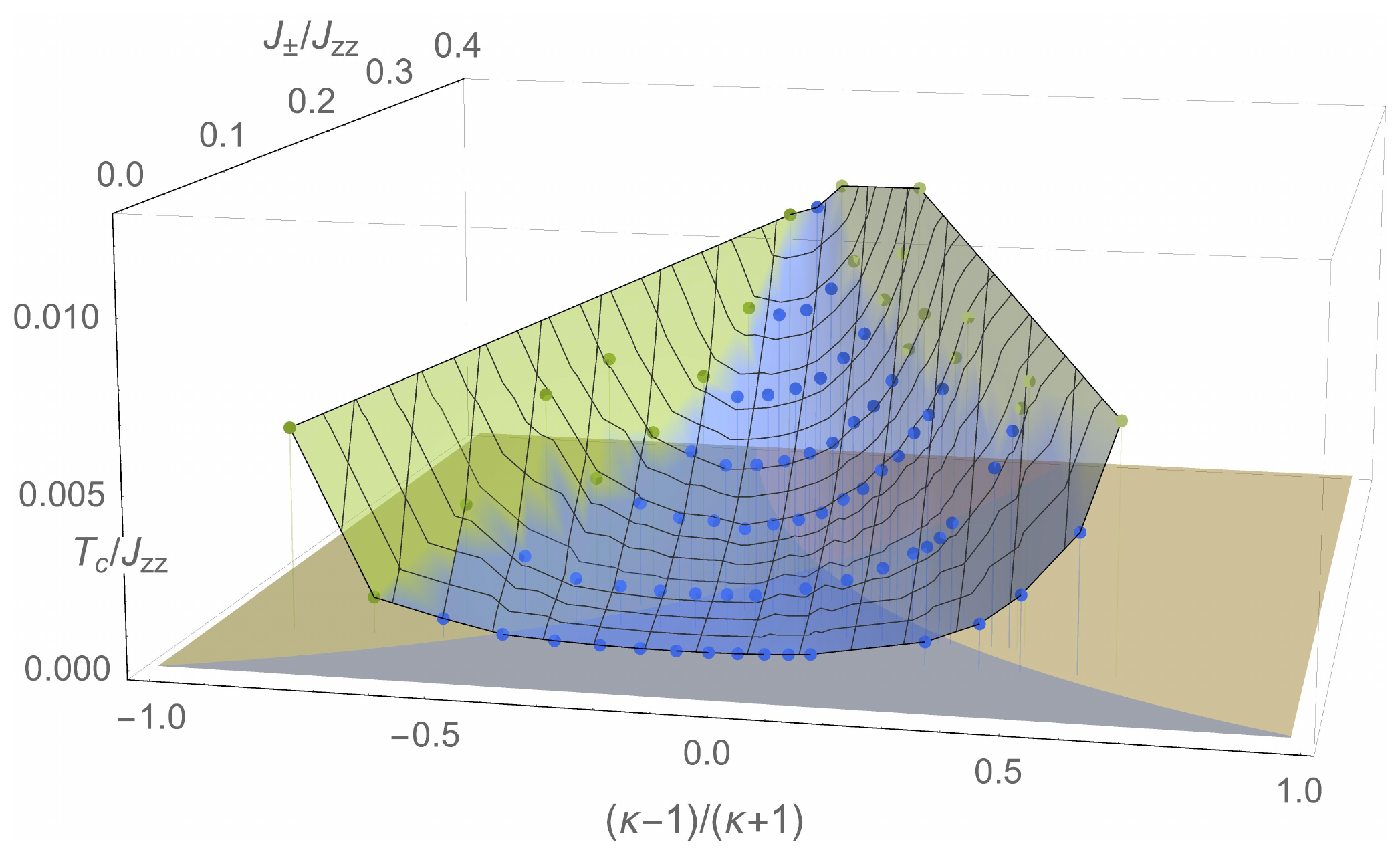} 
\caption{(Color online.) Phase diagram in the
  $\kappa$-$J_\pm/J_{zz}$-$T/J_{zz}$ space. 
  The three-dimensional surface indicates the first-order transition 
  between the phases akin to the zero-temperature phases, 
  namely the QSL, AFM, PM$_{\rm u,d}$ (below the surface), 
  and the TSL (above the surface). The colors on the
  surface correspond to those of the phases immediately 
  below the surface.}
\label{fig2}
\end{figure}

\section{Nonzero temperature spin liquid regime of QSI}
\label{sec4}

Now we turn to the nonzero temperature behavior of the system.
Strictly speaking, the QSI QSL is truly a quantum phase and 
cannot exist at non-zero temperatures.
This is because the topological defects (or magnetic monopole excitations) 
of the emergent $U(1)$ gauge field~\cite{Hermele04} 
in the QSI QSL state are point-like particles 
and have a finite energy~\cite{Savary13}. 
These topological excitations can always be 
created with a finite concentration at a non-zero temperature
so that the infinitesimal temperature phase above 
the QSL can be smoothly connected to the 
high-temperature phase~\cite{Savary13}. The above general argument
does not however preclude the existence of a non-generic first-order
transition, which indeed is found in nonzero temperature gMFT between
the low temperature regime akin to the QSL, and a classical regime where the 
entropy is released and the spinons lose coherence: the thermal spin liquid (TSL)
regime. This TSL regime is the one found in classical spin 
ices~\cite{GingrasSI,PhysRevLett.84.3430,RevModPhys.82.53}. 
Regardless of whether a crossover or a first order transition occurs
between the low temperature spin liquid regime and a thermal spin ice regime,
the temperature scale of either is set by the strength of the ring 
exchange~\cite{Hermele04,Savary13}, 
$J_{\text{ring}} = 6 (\kappa^3 + \kappa^{-2}) J_{\pm}^3/J_{zz}^2$. 
The minimum of $(\kappa^3 + \kappa^{-2})$, and hence that of $J_{\text{ring}}$, 
occurs at $\kappa \approx 0.92$ (see Appendix.~\ref{asec2}). 
We therefore expect that, at any given value of $J_{\pm}/J_{zz}$, the
crossover/transition temperature is usually enhanced for the breathing
pyrochlore lattice as compared to that in its regular counterpart. 
This suggests that the QSL physics can be observed at 
higher temperatures in breathing pyrochlore systems.

To explicitly confirm the above observation, we carry out a nonzero 
temperature gMFT calculation for the minimal model Eq.~(\ref{eq1}). The
calculation is carried out variationally, and the details are given in
Appendix.~\ref{asec4}. 
The gMFT finds a direct first-order transition, with the transition
temperature shown on the surface plotted on Fig.~\ref{fig2}. 
Just as we expect, the transition temperature is significantly enhanced 
as $\kappa$ deviates from 1. This transition can 
be regarded as a ``thermal confinement'' transition as 
discussed in early literature~\cite{PhysRevB.92.100403,Kato15,Savary13}.

Like many other mean field treatments, the gMFT neglects the 
gauge field fluctuations and thus overestimates the transition
temperature. This can be seen for example through the analytical gMFT
expression (see Appendix.~\ref{asec4}) obtained in the limit of very low temperature,
$\kappa\approx1$ and small $J_\pm/J_{zz}$:
\begin{equation}
k_B T_c^{\rm gMFT} =\frac{3J_\pm^2}{32J_{zz}\ln2 }\left(\kappa^2+\frac{1}{\kappa}\right)
\end{equation}
(to be compared, in that limit, with the $J_{\rm ring}$ scale of
perturbation theory).
Nevertheless, the gMFT does qualitatively capture the thermal activation of the
gapped spinon excitations which effectively reduces the magnetic stiffness and thus
destabilizes the QSL. 

\section{Discussion}
\label{sec5}

Since the featureless PM phases (PM$_{\text u}$ or PM$_{\text d}$) are also 
connected to the high-temperature paramagnet and do not have any ordering, 
one may then wonder how to distinguish the QSI QSL state and 
the featureless PM phases of Fig.~\ref{fig1}b 
in experiment. QSL has a linearly-dispersing gauge photon mode that 
gives rise to a low-temperature heat capacity $C_v \sim T^3$ whose 
prefactor can be quite large in 4f electron systems~\cite{Savary12,Shannon12,Chen13}. 
Inelastic neutron scattering and/or NMR spin lattice relaxation time 
measurements are direct probes of the gapless gauge photon. In contrast, 
since the PM$_{\text{u,d}}$ is fully gapped, 
the heat capacity should have an activated temperature dependence.  
 
Thermodynamic measurements on the breathing pyrochlore material
Ba$_3$Yb$_2$Zn$_5$O$_{11}$ have not found any indication of ordering 
down to 0.38K~\cite{Kimura14}. 
The magnetic entropy density extracted from the heat 
capacity data above 0.38K is close to $0.75 {\text R} \ln (2)$ per site, 
and the missing entropy ($0.25 {\text R} \ln (2) $ per site) was interpreted in
terms of an AFM Heisenberg model, effectively acting only within the smaller tetrahedra at
the accessible temperatures~\cite{Kimura14}: this model of decoupled tetrahedra has a
two-fold (the two singlet states of four coupled $S=1/2$) degeneracy
per small tetrahedron, giving, per site, an entropy of $0.25 {\text R} \ln (2) $.
In this picture, the exchange coupling on the large 
tetrahedra then operates perturbatively on the singlet manifold and lifts the 
degeneracy only at a much lower energy scale. (The
Heisenberg model for arbitrary small to large exchange was also
recently studied classically in detail in Ref.~\onlinecite{benton2015ground}.) 
Here we provide an alternative explanation for the experimental value
of the low-temperature magnetic entropy density. 

First, we argue that, due to strong spin-orbit coupling, the $J=7/2$ local moment 
of the Yb$^{3+}$ ion is an entangled state 
of the $L=3$ orbital angular momentum and $S=1/2$ spin moment, and an
isotropic model, such as the Heisenberg model, is therefore unlikely~\cite{William14,PhysRevB.78.094403,
PhysRevB.82.174440,*PhysRevB.84.094420}. This is especially true considering 
that the main contribution to the local doublet comes from the orbital degree of
freedom~\cite{Ross11}, which naturally singles out the local $z$ directions. 
This non-Heisenberg exchange, and more specifically its $D_{3d}$
anisotropy, points to the Pauling entropy of thermal spin ice
in the temperature regime $J_{\text{ring}} \lesssim T \lesssim \min(1,\kappa)J_{zz}$~\cite{Savary13}. 
Indeed, current experiments cannot rule out this possibility as 
Pauling's entropy ($0.5 {\text R} \ln (3/2)\approx0.20 {\text R}$ per site) is very close 
to the reported $0.25 {\text R} \ln (2) \approx0.17 {\text R}$ per site, with these
two values likely being within error bars of one another. 

At this point, neutron scattering experiments are crucially 
needed to check whether the system might be in the ice 
manifold and whether quantum spin ice QSL physics appears below $0.38$K. It will
also be interesting to look for QSL physics in other breathing pyrochlore systems with
other rare earth elements \cite{ZAAC:ZAAC19966220115}. 
Indeed, in general, based on our theoretical results, breathing pyrochlores
appear as a promising playground for the discovery of quantum spin ice
physics and novel quantum phase transitions, 
and certainly warrant further investigation.

\section{Acknowledgements} 

LS acknowledges Leon Balents for previous collaborations on related
work. LS is supported by a postdoctoral fellowship from the Gordon and Betty
Moore foundation, EPiQS initiative, Grant No.\ GBMF4303. HYK and YBK are 
supported by NSERC, Centre for Quantum Materials at UofT. 
GC would like to thank the hospitality of Fuchun Zhang and Yi Zhou 
for supporting his stay at Zhejiang University (Hangzhou, People's Republic of China) 
in April and May 2015 when and where part of the work was carried out. 
GC is supported by the Starting-up fund of Fudan University 
(Shanghai, People's Republic of China) and Thousand-Youth-Talent 
Program of People's Republic of China. Notes added: we recently noticed an
interesting work on a related topic~\cite{Rau}. 
The view of Ref.~\onlinecite{Rau} is complementary to ours.


\appendix

\section{Local coordinate system}
\label{asec1}

The Bravais lattice of the breathing pyrochlore lattice 
is a FCC lattice. We choose the basis vectors to be 
\begin{equation}
{\bf a}_1 = \frac{1}{2}[011], \; 
{\bf a}_2 = \frac{1}{2}[101],\; 
{\bf a}_3 = \frac{1}{2}[110],
\end{equation}
where the lattice constant is set to be unity. We define 
the reference points of the four sublattices as 
\begin{eqnarray}
{\bf b}_1 &=& [000],\\
{\bf b}_2 &=& c[011], \\
{\bf b}_3 &=& c[101], \\
{\bf b}_4 &=& c[110],
\end{eqnarray}
where $c=1/4$ for the regular pyrochlore lattice and 
$c \neq 1/4$ for the breathing pyrochlore lattice. The 
local coordinate system at each sublattice is then 
defined in Table~\ref{tab3}. 
\begin{table}[h]
\begin{tabular}{C{1cm}C{1.6cm}C{1.6cm}C{1.6cm}C{1.6cm}}
\hline\hline
$\mu$ & 1& 2& 3& 4 \\
$\hat{x}_{\mu}$ & $\frac{1}{\sqrt{2}} [\bar{1}10] $& $\frac{1}{\sqrt{2}}[\bar{1}\bar{1}0]$  &$\frac{1}{\sqrt{2}}[110]$  & $\frac{1}{\sqrt{2}}[1\bar{1}0]$  \\
$\hat{y}_{\mu}$ & $\frac{1}{\sqrt{6}}[\bar{1}\bar{1}2]$ & $\frac{1}{\sqrt{6}}[\bar{1}1\bar{2}] $ & $\frac{1}{\sqrt{6}}[1\bar{1}\bar{2}] $   &  $\frac{1}{\sqrt{6}}[112]  $\\
$\hat{z}_{\mu}$ & $\frac{1}{\sqrt{3}}[111]$ &  $\frac{1}{\sqrt{3}}[1\bar{1}\bar{1}] $ & 
$\frac{1}{\sqrt{3}}[\bar{1}1\bar{1}] $ & $\frac{1}{\sqrt{3}}[\bar{1}\bar{1}1] $ \\
\hline\hline
\end{tabular}
\caption{The local coordinate systems for the four sublattices of the breathing pyrochlore lattice. }
\label{tab3}
\end{table}

\section{Ring exchange}
\label{asec2}

\begin{figure}[tp]
\centering
 \includegraphics[width=6.5cm]{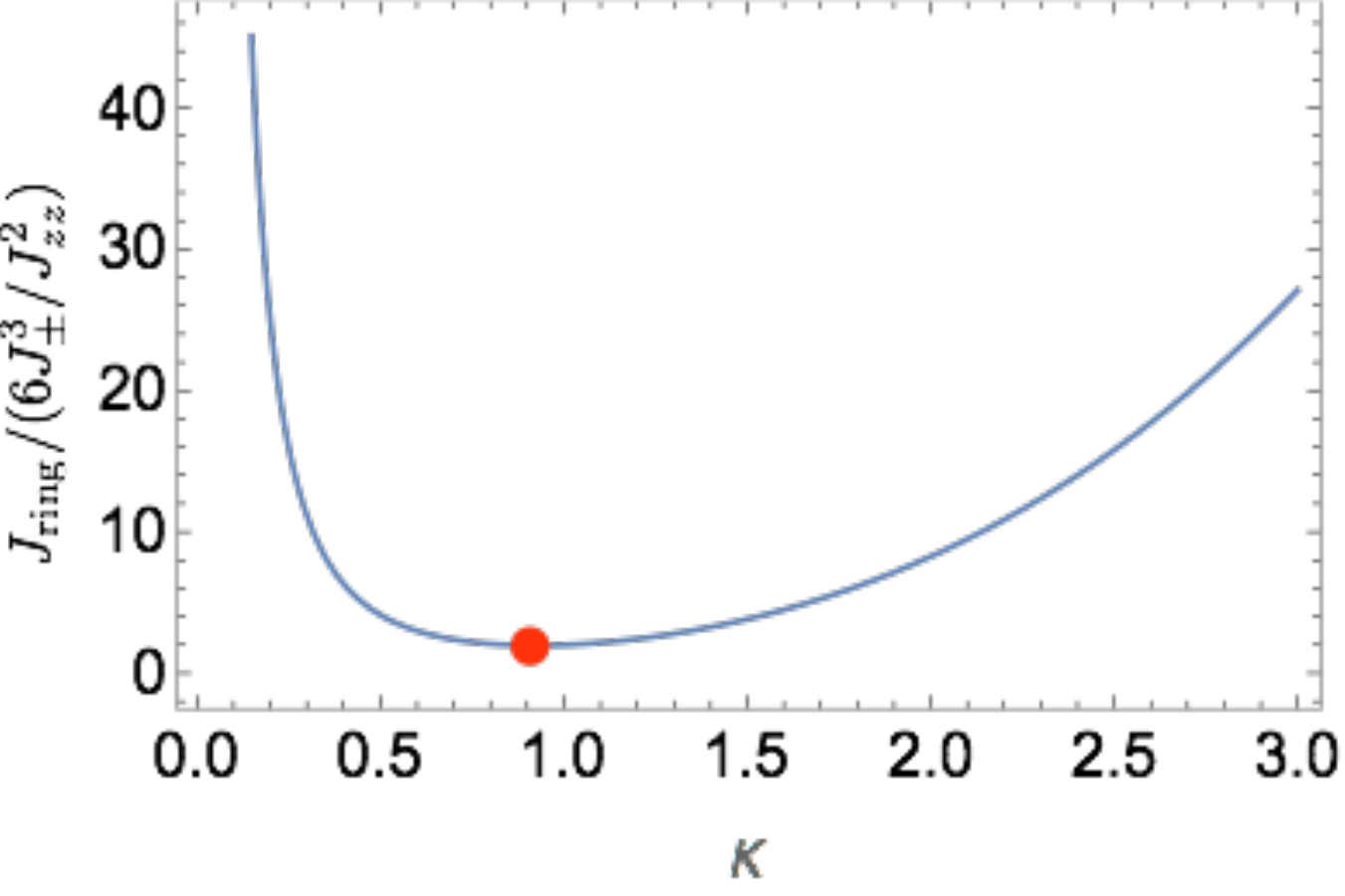} 
\caption{(Color online.) The ring exchange as 
a function of $\kappa$. The (red) dot is the position of 
the minimum and is at $\kappa \approx 0.92$. }
\label{fig3}
\end{figure}

In the perturbative regime or in the Ising limit, one can treat the $J_{\pm}$ 
and $\kappa J_{\pm}$ terms as a perturbation. We thus write the Hamiltonian 
$H$ as
\begin{equation}
H = H_{\text{Ising}} + H',
\end{equation} 
where $H_{\text{Ising}}$ and $H'$ are the unperturbed part and the perturbation part, respectively. We have 
\begin{eqnarray}
H_{\text{ring}} &=& \sum_{\langle ij \rangle \in \text{u} } J_{zz} \tau_i^z \tau_j^z
+ \sum_{\langle ij \rangle \in \text{d} } \kappa J_{zz} \tau_i^z \tau_j^z 
\\
H' &=& - \sum_{\langle ij \rangle \in \text{u} } {J_{\pm}} (\tau_i^+ \tau_j^- + h.c. )
\nonumber \\
&& - \sum_{\langle ij \rangle \in \text{d} } {\kappa J_{\pm}} (\tau_i^+ \tau_j^- + h.c. ).
\end{eqnarray}
The Ising exchange $H_{\text{Ising}}$ favors a highly 
degenerate spin ice ground state. 
The perturbation $H'$ acts on the ice manifold. Treating $H'$ 
in the 3rd order degenerate perturbation theory, 
we obtain a ring exchange effective Hamiltonian~\cite{Hermele04}, 
\begin{equation}
H_{\text{eff}} =  - \sum_{\hexagon} J_{\text{ring}}
( \tau_1^+\tau_2^-\tau_3^+ \tau_4^- \tau_5^+ \tau_6^- + h.c.),
\end{equation}
where 1,2,3,4,5,6 label the 6 pseudospins on the perimeter 
of an elementary hexagon on the breathing pyrochlore lattice
and 
\begin{equation}
J_{\text{ring}} = 6 (\kappa^3 + \kappa^{-2} ) \frac{J_{\pm}^3}{{J_{zz}^2} }.
\end{equation}
We map the effective Hamiltonian $H_{\text{eff}}$ to a 
$U(1)$ lattice gauge theory by expressing $\tau^z_i = E_{{\bf r}{\bf r}'}, 
\tau^{\pm}_i = e^{\pm i A_{{\bf r}{\bf r}'}}$, where ${\bf r}$ (${\bf r}'$) 
belongs to the u (d) sublattice of the diamond lattice 
formed by the centers of the tetrahedra and $i$
is the pyrochlore lattice site shared by the two neighbor 
tetrahedra at ${\bf r}$ and ${\bf r}'$ (see Fig.~\ref{fig1}a). 
$E$ and $A$ are the lattice electric field and the vector 
gauge potential defined on the diamond lattice and 
$E_{{\bf r}{\bf r}'} = -E_{{\bf r}'{\bf r}}, 
A_{{\bf r}{\bf r}'} = - A_{{\bf r}'{\bf r}}$~\cite{Hermele04}. 
In terms of the gauge field operators, the low energy ring exchange 
is recast as the following $U(1)$ gauge theory on the diamond lattice,
\begin{equation}
H_{\text{eff}} = - K \sum_{\varhexagon} \cos ( \nabla \times A )_{\varhexagon}
\end{equation}
where $\varhexagon$ refers to the elementary hexagons on the diamond 
lattice and the magnetic stiffness $K = 2 J_{\text{ring}}$ is plotted in 
Fig.~\ref{fig3} as a function of $\kappa$.

\section{Gauge mean field theory for the ground state}
\label{asec3}

In the gauge mean field theory, we decouple the Hamiltonian $H$ 
into the spinon and gauge sectors. The spinon
sector mean field Hamiltonian is written as 
\begin{eqnarray}
H_{\text u} &=& \sum_{{\bf r} \in \text{u}} \frac{J_{zz}}{2} Q_{\bf r}^2
 - \kappa J_{\pm} \langle \mathsf{s}^{\pm} \rangle^2
\sum_{{\bf r}\in \text{d}, \mu\neq\nu} \Phi^\dagger_{{\bf r} 
- {\bf e}_{\mu}}\Phi^{\phantom\dagger}_{{\bf r} -{\bf e}_{\nu}},
\\
H_{\text d} &=&
\sum_{{\bf r} \in \text{d}} \frac{\kappa J_{zz}}{2} Q_{\bf r}^2 
-  J_{\pm} \langle \mathsf{s}^{\pm} \rangle^2
\sum_{{\bf r}\in \text{u}, \mu\neq\nu} 
\Phi^\dagger_{{\bf r} + {\bf e}_{\mu}}\Phi^{\phantom\dagger}_{{\bf r} +{\bf e}_{\nu}} , 
\end{eqnarray}
where the u and d sublattices of the dual diamond lattice are effectively decoupled,
and $\langle \mathsf{s}^{\pm} \rangle \equiv \langle \mathsf{s}^{\pm}_{{\bf r} {\bf r}'} \rangle $. 
The gauge sector mean field Hamiltonian is simply a 
Zeeman-field-like term that couples to 
each $\mathsf{s}^{\pm}_{{\bf r}{\bf r}'}$. For the choice of the zero gauge 
flux mean field state, the gauge sector is thus trivially 
solved and has $\langle \mathsf{s}^{\pm} \rangle = 1/2$. 
The bosonic spinon sectors are finally solved by applying the 
standard coherent state path integral with a saddle point 
approximation~\cite{Savary12,Sungbin12,Chen13}.

\section{Gauge mean-field theory at nonzero temperature}
\label{asec4}
 
Here we proceed using a variational approach, and consider the trial
Hamiltonian $H^0=H_\Phi^0+H_{\mathsf{s}}^0$, with
\begin{eqnarray}
  \label{eq:2}
  H_\Phi^0&=&\frac{J_{\rm u}}{2}\sum_{\mathbf{r}\in{\rm u}}Q_{\mathbf{r}}^2-\sum_{\mathbf{r}\in{\rm
      d},\mu\neq\nu}t_{\mu\nu}'{}^{\rm
              u}\Phi^\dagger_{\mathbf{r}-\mathbf{e}_\mu}\Phi^{}_{\mathbf{r}-\mathbf{e}_\nu}\nonumber\\
&&+\frac{
    J_{\rm d}}{2}\sum_{\mathbf{r}\in{\rm d}}Q_{\mathbf{r}}^2-
  \sum_{\mathbf{r}\in{\rm u},\mu\neq\nu}t_{\mu\nu}'{}^{\rm d}\Phi^\dagger_{\mathbf{r}+\mathbf{e}_\mu}\Phi^{}_{\mathbf{r}+\mathbf{e}_\nu}
\end{eqnarray}
and
\begin{equation}
  \label{eq:3}
  H_{\mathsf{s}}^0=-\sum_{\mathbf{r}\in{\rm u}}\sum_\mu \vec{\mathsf{h}}_\mu(\mathbf{r})\cdot\vec{\mathsf{s}}_{\mathbf{r},\mathbf{r}+\mathbf{e}_\mu}.
\end{equation}
The chosen variational parameters are $J_{\rm u,d}$ and $t'_{\mu\nu}{}^{\rm u,d}$.


Now, we rewrite:
\begin{eqnarray}
  \label{eq:7}
  \frac{J_\eta}{2}\sum_{\mathbf{r}\in\eta}
  Q_\mathbf{r}\rightarrow \sum_{\mathbf{r}\in\eta}\left\{\frac{J_\eta}{2}\Pi_\mathbf{r}^\dagger\Pi^{}_\mathbf{r}+\lambda_\eta(\Phi_\mathbf{r}^\dagger\Phi^{}_\mathbf{r}-1)\right\},
\end{eqnarray}
where $\eta={\rm u,d}$. The trial spinon Hamiltonian then becomes:
\begin{eqnarray}
  \label{eq:8}
  H_\Phi^0&\rightarrow&\frac{J_{\rm u}}{2}\sum_{\mathbf{r}\in{\rm
      u}}\Pi_{\mathbf{r}}^\dagger\Pi^{}_\mathbf{r}
      \nonumber \\
&-&\sum_{\mathbf{r}\in{\rm
      d},\mu\neq\nu}t_{\mu\nu}'{}^{\rm
    u}\Phi^\dagger_{\mathbf{r}-\mathbf{e}_\mu}\Phi^{}_{\mathbf{r}-\mathbf{e}_\nu}+\lambda_{\rm
  u}\sum_{\mathbf{r}\in{\rm
   u}}\Phi^\dagger_\mathbf{r}\Phi^{}_\mathbf{r}
   \nonumber \\
&+&\frac{
    J_{\rm d}}{2}\sum_{\mathbf{r}\in{\rm
   d}}\Pi_{\mathbf{r}}^\dagger\Pi^{}_\mathbf{r}
   \nonumber \\
&-&
  \sum_{\mathbf{r}\in{\rm u},\mu\neq\nu}t_{\mu\nu}'{}^{\rm d}\Phi^\dagger_{\mathbf{r}+\mathbf{e}_\mu}\Phi^{}_{\mathbf{r}+\mathbf{e}_\nu}+\lambda_{\rm
  d}\sum_{\mathbf{r}\in{\rm d}}\Phi^\dagger_\mathbf{r}\Phi^{}_\mathbf{r}.
\end{eqnarray}

Going to Fourier space and Matsubara frequency, and setting $t'_{\mu\nu}{}^{\rm u,d}=t_{\rm u,d}$ we obtain the Green's function for the trial
spinon Hamiltonian:
\begin{equation}
  \label{eq:9}
  {[G_0^{-1}]}=\left(\begin{array}{cc}
\frac{1}{2J_{\rm u}}\omega_n^2+\lambda_{\rm u}-\tilde{L}^{\rm u}_{\mathbf{k}} &
0\\
0 & \frac{1}{2J_{\rm d}}\omega_n^2+\lambda_{\rm d}-\tilde{L}^{\rm d}_{\mathbf{k}}
\end{array}\right),
\end{equation}
where
\begin{equation}
  \label{eq:10}
  \tilde{L}_\mathbf{k}^\eta=t_{\eta}\sum_{\mu\neq\nu}e^{i\mathbf{k}\cdot(\mathbf{e}_\mu-\mathbf{e}_\nu)}=t_\eta\sum_{\mu\neq\nu}\cos [\mathbf{k}\cdot(\mathbf{e}_\mu-\mathbf{e}_\nu)],
\end{equation}
and from which we may extract the spinon dispersion relations:
\begin{eqnarray}
  \label{eq:11}
  \omega^{\rm u}_\mathbf{k}&=&\sqrt{2J_{zz}}\sqrt{\lambda_{\rm
      u}-\tilde{L}_\mathbf{k}^{\rm u}},\\
\omega^{\rm d}_\mathbf{k}&=&\sqrt{2\kappa J_{zz}}\sqrt{\lambda_{\rm
      d}-\tilde{L}_\mathbf{k}^{\rm d}}.
\end{eqnarray}
Inverting $[G^{-1}_0]$ simply yields:
\begin{equation}
  \label{eq:13}
   [G_0]=\left(\begin{array}{cc}
\frac{1}{\frac{1}{2J_{\rm u}}\omega_n^2+\lambda_{\rm u}-\tilde{L}^{\rm u}_\mathbf{k}} &
0\\
0 & \frac{1}{\frac{1}{2J_{\rm d}}\omega_n^2+\lambda_{\rm
    d}-\tilde{L}^{\rm d}_\mathbf{k}}
\end{array}\right),
\end{equation}
and the equal-time version is obtained by summing over all
$\Omega_n=2\pi n/\beta$ ($\beta=1/(k_{\rm B}T)$ where $T$ is the
temperature and $k_{\rm B}$ Boltzmann constant), $n\in\mathbb{Z}$:
\begin{eqnarray}
  \label{eq:14}
  G_0(\tau=0)&=&\frac{1}{\beta}\sum_{n\in\mathbb{Z}}G_0(\mathbf{k},\Omega_n)\\
&=&\left(\begin{array}{cc}
\sqrt{\frac{J_{\rm u}}{2}}\frac{\mathcal{F}^{\rm u}_\mathbf{k}}{\sqrt{\lambda_{\rm u}-\tilde{L}^{\rm u}_\mathbf{k}}} &
0\\
0 & \sqrt{\frac{J_{\rm d}}{2}}\frac{\mathcal{F}^{\rm d}_\mathbf{k}}{\sqrt{\lambda_{\rm
    d}-\tilde{L}^{\rm d}_\mathbf{k}}}
\end{array}\right)
\end{eqnarray}
where
\begin{eqnarray}
  \label{eq:6}
\mathcal{F}^{\rm
  u}_\mathbf{k}&=&\coth\left[\beta\sqrt{\frac{J_{zz}}{2}}\sqrt{\lambda_{\rm
      u}-\tilde{L}_\mathbf{k}^{\rm u}}\right],\\
\mathcal{F}^{\rm d}_\mathbf{k}&=&\coth\left[\beta\sqrt{\frac{\kappa J_{zz}}{2}}\sqrt{\lambda_{\rm
      d}-\tilde{L}_\mathbf{k}^{\rm d}}\right].
\end{eqnarray}
Finally, if we choose
\begin{eqnarray}
  \label{eq:17}
J_{\rm u}&=&J_{zz} ,\\
J_{\rm d}&=&\kappa J_{zz} ,\\
  t^{\rm u}&=&\kappa J_\pm\mathsf{s}^2, \\
t^{\rm d}&=&J_\pm\mathsf{s}^2,
\end{eqnarray}
where $0\leq\mathsf{s}\leq1/2$ and ``represents'' $|\langle\vec{\mathsf{s}}\rangle|$, 
we need
to minimize 
\begin{eqnarray}
  \label{eq:5}
  \frac{F_v}{N_{\rm
    u.c.}}&=&4k_BT\left[(\frac{1}{2}+\mathsf{s})\ln(\frac{1}{2}+\mathsf{s})+(\frac{1}{2}-\mathsf{s})\ln(\frac{1}{2}-\mathsf{s})\right]\nonumber\\
&&-\lambda_{\rm
  u}-\lambda_{\rm d}\nonumber\\
&&+\frac{1}{N_{\rm u.c.}}\sum_{\mathbf{k}}\sum_{\eta={\rm
   u/d}}\left[\omega_\mathbf{k}^\eta-2k_BT\ln\frac{1}{1-e^{-\beta\omega_{\mathbf{k}}^\eta}}\right],\nonumber\\
\label{eq:5c}
\end{eqnarray}
while imposing the two rotor constraints (on average) simultaneously:
\begin{eqnarray}
  \label{eq:12}
  1=I_3^{\rm u}=\langle\Phi^\dagger_\mathbf{r}\Phi^{}_\mathbf{r}\rangle&=&\frac{1}{N_{\rm
      u.c.}}    \sum_\mathbf{k}\frac{\mathcal{F}^{\rm
    u}_\mathbf{k}  ({J_{zz}}/{2})^{\frac{1}{2}  } }{  [{\lambda_{\rm u}-\tilde{L}_\mathbf{k}^{\rm
      u}}]^{\frac{1}{2}}   },  \\
  1=I_3^{\rm d}=\langle\Phi^\dagger_\mathbf{r}\Phi^{}_\mathbf{r}\rangle&=&\frac{1}{N_{\rm
      u.c.}} \sum_\mathbf{k}\frac{\mathcal{F}^{\rm
    d}_\mathbf{k}  ( {\kappa J_{zz}}/{2})^{\frac{1}{2}  }  }{  [{\lambda_{\rm d}-\tilde{L}_\mathbf{k}^{\rm
      d}} ]^{\frac{1}{2}}  }.
\end{eqnarray}
The deconfined and condensed phases are distinguished in particular by
the existence of an important subextensive part to $\lambda^\eta$, which we write:
\begin{eqnarray}
  \label{eq:16}
  \lambda^{\rm u}&=&\lambda_{\rm min}^{\rm u}+ \frac{ \delta_{\rm u} \kappa T }{N_{\rm u.c.}}, \\
\lambda^{\rm d}&=&\lambda_{\rm min}^{\rm d}+  \frac{ \delta_{\rm d}  T}{N_{\rm
                   u.c.}} .
\label{eq:16b}
\end{eqnarray}
There, $\delta_\eta=O(1)$, $\delta_\eta\geq0$ and independent of temperature,
and
\begin{equation}
  \label{eq:15}
  \lambda_{\rm min}^\eta=\tilde{L}_{\mathbf{k}_0^\eta}^{\rm \eta}=\max_\mathbf{k}\tilde{L}^\eta_\mathbf{k}.
\end{equation}
$\mathbf{k}_0^\eta$ are the wavevectors at which the spinon dispersion
relations become gapless.

Defining
\begin{eqnarray}
  \label{eq:19}
  I_3^\eta{}'&=&\sqrt{\frac{J_\eta}{2}}\int_\mathbf{k}\frac{\mathcal{F}_\mathbf{k}^\eta}{\sqrt{\lambda_{\rm
      min}^\eta-\tilde{L}_\mathbf{k}^\eta}}
\end{eqnarray}
and 
\begin{equation}
  \label{eq:1}
  I_3^{\eta,{\rm min}}=\lim_{N_{u.c.}\rightarrow\infty}\sqrt{\frac{J_\eta}{2}}\frac{\mathcal{F}_{\mathbf{k}_0}^\eta}{\sqrt{\lambda^\eta-\tilde{L}_{\mathbf{k}_0}^\eta}}
\end{equation}
we have:
\begin{eqnarray}
  \label{eq:18}
  I_3^{\eta}=I_3^{\eta,{\rm min}}+I_3^{\eta}{}'.
\end{eqnarray}

\vspace{0.5cm}

\noindent{\bf Special values}\\
\noindent{The} phase diagram is obtained by minimizing Eq.~(\ref{eq:5c})
subjected to the constraints Eqs.~(\ref{eq:16},\ref{eq:16b}), and
following Table~\ref{tab2}. In general, the solution must be found
numerically. However, in some limits, thanks to some special values of
$\mathsf{s}$, it is possible to obtain analytical results for the
transition temperature. This is what is investigated below.

\vspace{0.4cm}
\noindent{\em $\mathsf{s}=0$ --- Thermal Spin Liquid}

If $\mathsf{s}=0$, which corresponds to the Thermal Spin Liquid state, then
\begin{eqnarray}
  \label{eq:20}
&&  \frac{F_v(\mathsf{s}=0)}{N_{\rm u.c.}}=\nonumber \\
&&4k_BT(-\ln2)-\lambda_{\rm
    u}-\lambda_{\rm d}+\bigg[\sqrt{2J_{zz}}\left(\sqrt{\lambda_{\rm
        u}}+\sqrt{\kappa\lambda_{\rm
        d}}\right)\nonumber\\
&&\left.-2k_BT\left(\ln\frac{1}{1-e^{-\beta\sqrt{2J_{zz}}\sqrt{\lambda_{\rm
        u}}}}+\ln\frac{1}{1-e^{-\beta\sqrt{2J_{zz}}\sqrt{\kappa\lambda_{\rm
        d}}}}\right)\right]\nonumber\\
\end{eqnarray}
and
\begin{eqnarray}
  \label{eq:21}
    1&=&\sqrt{\frac{J_{zz}}{2}}\frac{\coth\left[\beta\sqrt{\frac{J_{zz}}{2}}\sqrt{\lambda_{\rm
      u}}\right]}{\sqrt{\lambda_{\rm u}}},\\
  1&=&\sqrt{\frac{\kappa
      J_{zz}}{2}}\frac{\coth\left[\beta\sqrt{\frac{\kappa J_{zz}}{2}}\sqrt{\lambda_{\rm
      d}}\right]}{\sqrt{\lambda_{\rm d}}}.
\end{eqnarray}
For small enough temperature and small enough $T/\kappa$ (numerically,
the transition is indeed found to happen at small $T$) we find
\begin{equation}
  \label{eq:22}
 \lambda_{\rm u}=\frac{J_{zz}}{2},\qquad
 \lambda_{\rm d}=\frac{\kappa J_{zz}}{2},
\end{equation}
and therefore:
\begin{eqnarray}
  \label{eq:23}
 &&   \frac{F_v(\mathsf{s}=0)}{N_{\rm u.c.}}
 \nonumber \\
&&=-2k_BT\left(2\ln2+\ln\frac{1}{1-e^{-\beta
      J_{zz}}}+\ln\frac{1}{1-e^{-\beta\kappa
      J_{zz}}}\right)\nonumber\\
&&\quad+\frac{J_{zz}}{2}\left(1-\kappa+2\sqrt{\kappa}\right)\\
&&\approx-4k_BT\ln2+\frac{J_{zz}}{2}\left(1-\kappa+2\sqrt{\kappa}\right)
  \label{eq:23b}
\end{eqnarray}
for $T$ and $T/\kappa$ small enough. Note that we recover the
$\kappa=1$ result at $\kappa=1$.

\vspace{0.4cm}
\noindent{\em $\mathsf{s}=1/2$ --- Zero temperature limit}

When $\mathsf{s}=1/2$, which corresponds to zero temperature (but,
numerically, at small temperature, and below the transition,
$\mathsf{s}$ deviates only very slightly from $1/2$) and
for small enough temperature (i.e.\ in particular if $\lambda_\eta(T)\approx\lambda_\eta(T=0)$),
\begin{eqnarray}
  \label{eq:24}
&&  \frac{F_v(\mathsf{s}=1/2)}{N_{\rm u.c.}}
\nonumber \\
&&\approx  \frac{F_v(T=0)}{N_{\rm
    u.c.}}-2k_BT\sum_{\eta={\rm
      u/d}}\int_\mathbf{q}\ln\frac{1}{1-e^{-\beta\omega_\mathbf{q}^\eta}}
      \nonumber \\
&&\approx \frac{F_v(T=0)}{N_{\rm
    u.c.}}+2k_BT\sum_{\eta={\rm
      u/d}}\int_\mathbf{q}\ln[\beta\omega_\mathbf{q}^\eta]
      \nonumber \\
&&\approx \frac{F_v(T=0)}{N_{\rm u.c.}}, 
\end{eqnarray}
and 
\begin{eqnarray}
  \label{eq:25}
 && \frac{F_v(T=0)}{N_{\rm u.c.}}\approx-\lambda_{\rm u}-\lambda_{\rm
    d}\nonumber
    \\
&&+\sqrt{2J_{zz}}\int_\mathbf{q}\left(\sqrt{\lambda_{\rm
      u}-\frac{1}{4}\kappa J_\pm\hat{L}_\mathbf{q}}+\sqrt{\kappa\lambda_{\rm
      d}-\frac{1}{4}\kappa J_\pm\hat{L}_\mathbf{q}}\right).\nonumber \\
\end{eqnarray}
Now, for small $J_\pm$,
\begin{eqnarray}
  \label{eq:26}
 &&   \frac{F_v(T=0)}{N_{\rm u.c.}} \nonumber \\
&&\approx-\lambda_{\rm u}-\lambda_{\rm
    d} \nonumber \\
&&+\sqrt{2J_{zz}}\int_\mathbf{q}\left[\sqrt{\lambda_{\rm
        u}}\left(1-\frac{1}{8}\frac{\kappa J_\pm}{\lambda_{\rm
   u}}\hat{L}_\mathbf{q}-\frac{1}{128}\frac{\kappa^2
   J_\pm^2}{\lambda_{\rm u}^2}\hat{L}^2_\mathbf{q}\right)\right. \nonumber \\
&&\left.+\sqrt{\kappa\lambda_{\rm
      d}}\left(1-\frac{1}{8}\frac{ J_\pm}{\lambda_{\rm
        d}}\hat{L}_\mathbf{q}-\frac{1}{128}\frac{ J_\pm^2}{\lambda_{\rm
        d}^2}\hat{L}^2_\mathbf{q}\right)\right] \nonumber \\
&&\approx-\lambda_{\rm u}-\lambda_{\rm
    d}\nonumber \\
&&+\sqrt{2J_{zz}}\left[\sqrt{\lambda_{\rm
        u}}+\sqrt{\kappa\lambda_{\rm d}}-\frac{3 J_\pm^2}{32}\left(\frac{\kappa^2}{\lambda_{\rm u}^{3/2}}+\frac{\sqrt{\kappa}}{\lambda_{\rm
        d}^{3/2}}\right)\right].
\nonumber        
        \\
\end{eqnarray}
At $T=0$, the $\lambda$ constraints are
\begin{eqnarray}
  \label{eq:27}
    1&=&\frac{1}{N_{\rm
      u.c.}}\sqrt{\frac{J_{zz}}{2}}\sum_\mathbf{k}\frac{1}{\sqrt{\lambda_{\rm
        u}-\frac{\kappa J_{\pm}}{4}\hat{L}_\mathbf{k}}}, \\
  1&=&\frac{1}{N_{\rm
      u.c.}}\sqrt{\frac{\kappa J_{zz}}{2}}\sum_\mathbf{k}\frac{1}{\sqrt{\lambda_{\rm d}-\frac{J_\pm}{4}\hat{L}_\mathbf{k}}},
\end{eqnarray}
which is, for small enough $J_\pm$,
\begin{eqnarray}
  \label{eq:28}
  1&\approx&\frac{1}{N_{\rm
      u.c.}} (\frac{J_{zz}}{2 \lambda_{\rm
        u}})^{\frac{1}{2}}  
   \sum_\mathbf{k} [1+\frac{\kappa J_{\pm}}{8\lambda_{\rm u}}\hat{L}_\mathbf{k}+\frac{3\kappa^2 J_{\pm}^2}{128\lambda_{\rm u}}\hat{L}_\mathbf{k}^2 ], \\
  1&\approx&\frac{1}{N_{\rm
      u.c.}}   (\frac{J_{zz}}{2 \lambda_{\rm
        d}})^{\frac{1}{2}}    
      \sum_\mathbf{k} [1+\frac{J_{\pm}}{8\lambda_{\rm u}}\hat{L}_\mathbf{k}+\frac{3 J_{\pm}^2}{128\lambda_{\rm d}}\hat{L}_\mathbf{k}^2] ,
\end{eqnarray}
and finally
\begin{eqnarray}
  \label{eq:29}
  1&=&\sqrt{\frac{J_{zz}}{2 \lambda_{\rm
        u}}}\left(1+\frac{3\kappa^2 J_{\pm}^2}{32\lambda_{\rm u}}\right), \\
  1&=&\sqrt{\frac{\kappa J_{zz}}{2\lambda_{\rm d}}}\sum_\mathbf{k}\left(1+\frac{3 J_{\pm}^2}{32\lambda_{\rm d}}\right),
\end{eqnarray}
which again yields, as a first approximation,
\begin{equation}
  \label{eq:30}
  \lambda_{\rm u}\approx\frac{J_{zz}}{2}\qquad\mbox{and}\qquad \lambda_{\rm d}\approx\frac{J_{zz}\kappa}{2},
\end{equation}
and so
\begin{eqnarray}
  \label{eq:31}
  \frac{F_v(T=0)}{N_{\rm
      u.c.}}&\approx&-\frac{J_{zz}}{2}\left(1+\kappa\right) \nonumber \\
&&+J_{zz}\left[1+\sqrt{\kappa}-\frac{3
      J_\pm^2}{8
      J_{zz}^2}\left(\kappa^2+\frac{\sqrt{\kappa}}{\kappa^{3/2}}\right)\right]\nonumber\\
&\approx&\frac{J_{zz}}{2}\left[1-\kappa+2\sqrt{\kappa}-\frac{3
      J_\pm^2}{4
      J_{zz}^2}\left(\kappa^2+\frac{1}{\kappa}\right)\right].
      \nonumber \\
  \label{eq:31b}
\end{eqnarray}
Again, at $\kappa=1$, we recover the `regular pyrochlore' result.

The transition between a regime akin to the Coulombic QSL 
(with $\mathsf{s}\lesssim1/2$) and the
Thermal Spin Liquid (with $\mathsf{s}=0$) is found numerically to be strongly first-order
and to occur at low temperature. So, equating Eqs.~(\ref{eq:23b}) and (\ref{eq:31b}), we obtain:
\begin{equation}
  \label{eq:32}
  T_c\approx\frac{3J_\pm^2}{32J_{zz}k_B\ln2 }\left(\kappa^2+\frac{1}{\kappa}\right),
\end{equation}
which we obtained assuming small $T$, small $T/\kappa$ and small $J_\pm$.

\bibliography{ref}

\end{document}